# RF Positive Ion Source with Solenoidal Magnetic Field

V. Dudnikov[1,a], R.P. Johnson[1], B. Han[2], Y. Kang[2], S. Murray[2],
T. Pennisi[2], C. Piller[2], C. Stinson[2], M. Stockli[2], R. Welton[2],
G. Dudnikova[3,4]

[1]*Muons, Inc., Batavia, IL 60510, USA;* [2]*ORNL, Oak Ridge, TN 37831, USA;*
[3]*University of Maryland, College Park, MD- 32611-USA;*
[4]*Institute of Computational Technologies SBRAS, Novosibirsk, Russia*
[a]Corresponding author: Vadim@muonsinc.com

**Abstract.** A positive ion source with RF discharge in solenoidal magnetic field is described. In this paper we present an overview of positive ion production in saddle antenna (helicon discharge) radio frequency (SA RF) ion sources. The efficiency of H+ ion production in recently developed RF sources with solenoidal antennas was improved to 2.9 mA/kW. About 24 kW of RF power is typically needed for 70 mA beam current production from a 7 mm emission aperture. This efficiency is relatively low because in the RF discharge with a solenoidal antenna, the plasma is generated near the coil and diffuses to the axis creating a nearly uniform plasma density distribution in all cross sections of the discharge chamber, when the plasma flow is necessary only near an emission aperture. The efficiency of the extracted ion generation was improved significantly by using a saddle antenna with solenoidal magnetic field. In the RF discharge with the saddle antenna the plasma is generated near the axis and the magnetic field suppress the plasma diffusion from the axis, creating a peaked plasma density distribution on the emission aperture. With the SA the efficiency of positive ion generation in the plasma has been improved up to ~100 mA/cm$^2$ per kW of RF power at 13.56 MHz. Continuous wave (CW) operation of the RF source has been tested on the small ORNL SNS test stand. The general design of the CW RF source is based on the pulsed version. A compact design of ion source is presented. Some modifications were made to improve the cooling and to simplify the design. Features of SA RF discharges and ion generation are discussed.

## INTRODUCTION

The efficiency of H+ ion production in recently developed RF sources with solenoidal antennas was improved to 2.9 mA/kW [1]. About 24 kW of RF power is typically needed for 70 mA beam current production from a 7 mm emission aperture. This efficiency is relative low because in the RF discharge with solenoidal antenna, the plasma is generated near the coil and diffuses to the axis creating a nearly uniform plasma density distribution in all cross sections of the discharge chamber, when the plasma flow is necessary only near an emission aperture. The efficiency of the extracted ion generation was improved significantly by using of the saddle antenna and solenoidal magnetic field [2-10]. In the RF discharge with the saddle antenna the plasma is generated near the axis and the magnetic field suppresses the plasma diffusion from the axis, creating a peaked plasma density distribution on the emission aperture. With the SA the efficiency of positive ion generation in the plasma has been improved up to 100 mA/cm$^2$ per kW of RF power at a frequency of 13.56 MHz.

## RF ION SOURCE IN THE SNS TEST STAND

Figure 1(a) is the schematic of the large RF ion source, showing the AlN ceramic discharge chamber, saddle antenna, and DC 50 turn solenoid, up to 70 A. The chamber has an ID=68 mm. The saddle antenna with inductance L=3.5 μH is made from a water-cooled copper tube. The RF assisted triggering plasma gun (TPG) is attached to the discharge chamber on the left [4, 7]. The extraction system is attached on the right side. The plasma in the TPG is generated by a continuous wave (CW) RF discharge (13.56 MHz, ~ 250 W) and electrons are injected into main discharge chamber by the extraction voltage. The ion source is inserted into a vacuum chamber pumped by a turbo molecular pump. The upper window is used for beam extraction observation. The right-hand side window is used to observe the back side of the collector while it is heated by ion beams.

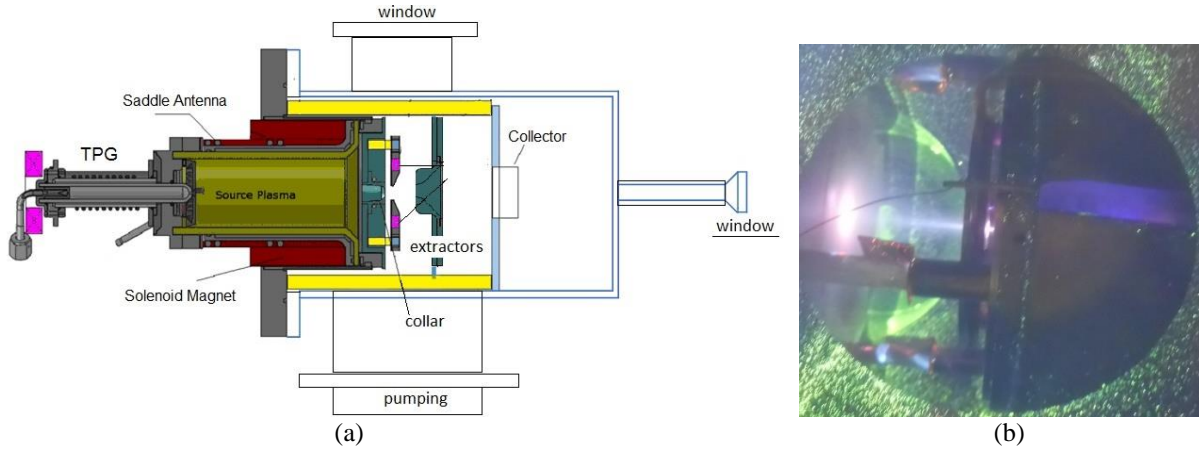

(a)            (b)

**FIGURE 1**. (a) A schematic of the SA ion source with an extraction system, collector in vacuum chamber; (b) Photograph of formation of the positive ion beam. The collector is heated up to high temperature.

The plasma flux, generated in the AlN discharge chamber by the saddle antenna is guided by the longitudinal magnetic field (created by the DC solenoid) to a plasma electrode with a conical collar that defines the emission aperture of 6 mm diameter. Ions are extracted from the cone aperture 6mm by the extraction voltage Uex between the cone and the extractor attached to the plasma electrode through ceramic insulators, and accelerated by the voltage across the second gap by voltage Uc on the accelerator electrode. Beam is damped to a collector with suppression of secondary emission.

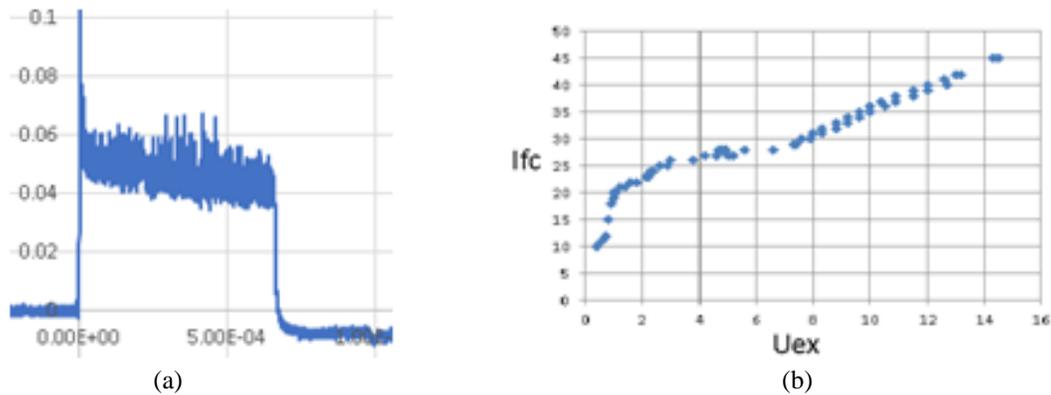

(a)            (b)

**FIGURE 2**. (a) Signals of positive ion on collector, Ic = 50 mA at RF power ~1.5 kW in the plasma. The time scale is 0.5 ms/div;( b) Dependence of collector current Ic (mA) on extraction voltage (kV), Prf=1.5 kW, solenoid voltage Um=5 V.

The plasma flux is compressed by the increase of solenoidal magnetic field. The evidence of this plasma flux behavior is the trace of the dark film deposited on the conical collar surface. With the increase of the solenoid current, this deposited area decreases up to the emission aperture. The SA ion source was tested with 6 mm diameter emission and extractor apertures. A photograph of proton beam extraction is shown in Fig. 1(b). It is visible light from the positive ion beam and the collector, heated to high temperature. The oscilloscope trace of the collector current of 50 mA is shown in Fig. 2(a). The dependence of the collector current Ic on extraction voltage is shown in Fig. 2(b). It is possible to have Ic ~ 50 mA at extraction voltage 16 kV.

For positive ion extraction, a simplified version of the RF ion source was prepared with the solenoidal magnetic field as shown in Fig. 3. In this version the extractor electrode was made thinner. In CW mode the discharge can be triggered at high gas pressure and after the discharge the triggering gas pressure can be decreased. In this case we don't need the triggering plasma gun.

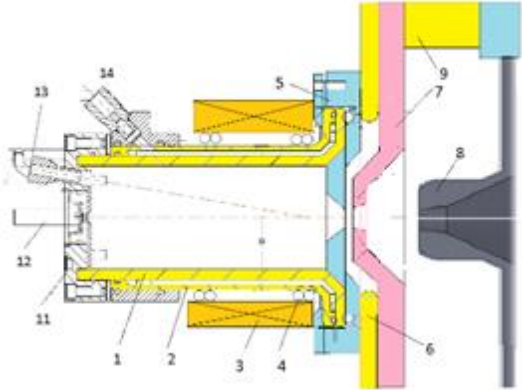

**FIGURE 3.** Schematic of RF discharge positive ion source with solenoidal magnetic field.

1-Gas discharge chamber (AlN), 2- cooling jacket from KEEP, 3- solenoid, 4- saddle antenna, 5-plasma electrode with conical collar and emission aperture, 6-extractor insulator, 7- extraction electrode, 8- grounded electrode, 9- insulator, 11- back flange, 12- gas inlet, 13- view port, 14- cooling water inlet-outlet.

CW operation of the SA positive ion source with positive ion extraction was tested with RF power up to ~ 2 kW from the generator (~ 1.5 kW in the plasma, and 0.5 kW is dissipated in antenna, network and solenoid) with production up to Ic=50 mA for short time. Long term operation was tested with 1.8 kW from the RF generator (~ 1.3 kW in the plasma) with production of Ic=45 mA (Uex=15 kV). A 30 days test in CW mode was successfully formed with no damage to the source.

The collector current is increase with increase of a magnetic field up to solenoid voltage Um ~8 V. The specific power efficiency of positive ion beam production in CW mode is up to Spe ~ 100 mA/cm$^2$ kW.

CW RF discharge can be triggered with CW discharge in the Triggering Plasma Gun (TPG) at gas flow Q ~ 8 sccm and can be supported up to Q ~ 3 sccm. The main CW discharge in SA RF SPS can be triggered without discharge in the TPG at Q ~ 10 sccm and supported up to Q ~ 4 sccm.

## ACKNOWLEDGEMENTS

The work was supported in part by US DOE Contract DE-AC05-00OR22725 and by STTR grant DE-SC0011323.

## REFERENCES

1. J. Lettry, D. Aguglia, S. Bertolo, et al., "CERN's Linac4 cesiated surface H- source," AIP Conference Proceedings 1869 (2017), p.030002.
2. V. Dudnikov et al., AIP Conference Proceedings 925 (2007), pp.153-163.
3. V. Dudnikov et al., in THPS026, Proceedings of IPAC2011, San Sebastian, Spain, 2011.
4. V. Dudnikov et al., Rev. Sci. Instrum. **83**, 02A712 (2012).
5. V. Dudnikov et al., AIP Conference Proceedings 1390 (2011), p. 411.
6. V. Dudnikov et al., AIP Conference Proceedings 1515 (2013), p.456.
7. V. Dudnikov et al., AIP Conference Proceedings 1655 (2015), p. 070003
8. R. F. Welton et al., Rev. Sci. Instrum. **83**, 02A725 (2012).
9. V. Dudnikov, A. Dudnikov, Rev. Sci. Instrum. **83** (2), 02A720 (2012).
10. V. Dudnikov, P. Chapovsky, and A. Dudnikov, Rev. Sci. Instrum. **81**, 02A714 (2010).